\documentclass[9pt, twocolumn, twoside=false, head=13pt,
     paperwidth=11in, paperheight=8.5in,
     includeheadfoot, marginparsep=72pt,
     marginparwidth=170pt, columnsep=20pt,
     top=72pt, bottom=72pt, left=314pt, right=72pt]{article}%

\usepackage{booktabs} 

\usepackage{graphicx}  
\usepackage{subcaption}
\usepackage{amsmath}%
\usepackage{amsfonts}%
\usepackage{amssymb}%
\usepackage{graphicx}

\begin{document}
\title{Phase Transition in the Maximal Influence Problem:\\When Do We Need Optimization?}
 
\author{
Yoav Kolumbus \and
Sorin Solomon
}
 
\date{
Racah Institute of Physics, the Hebrew University of Jerusalem, Israel\\[2ex]
}
\maketitle

\begin{abstract}
Considerable efforts were made in recent years in devising optimization algorithms for influence maximization in networks. Here we ask: ``When do we need optimization?'' We use results from statistical mechanics and direct simulations on ER networks, small-world networks, power-law networks and a dataset of real-world networks to characterize the parameter-space region where optimization is required. We show that in both synthetic and real-world networks this optimization region is due to a well known physical phase transition of the network, and that it vanishes as a power-law with the network size. We then show that also from a utility-maximization perspective (when considering the costs of the optimization process), for large networks standard optimization is profitable only in a vanishing parameter region near the phase transition. Finally, we introduce a novel constant-time optimization approach, and demonstrate it through a simple algorithm that manages to give similar results to standard optimization methods in terms of the influenced-set size, while improving the results in terms of the net utility.
\end{abstract}

%
%
%


\maketitle

\section{Introduction} \label{sec:intro}
The present work discusses the process of diffusion in networks, which has been widely studied in various contexts, including sociology  \cite{strang1998diffusion,ryan1943diffusion, rogers2010diffusion}, marketing  \cite{Goldenberg2001,brown1987social,Domingos:2001:MNV:502512.502525,peres2010innovation}, economics  \cite{cowan2004network,jackson2007diffusion, banerjee2013diffusion}, and physics  \cite{callaway2000network, 
PhysRevE.62.7059, albert2002statistical}. 
We focus on the common setting in the context of optimization, in which the 
process starts from an initial seed-set of influencers, and then spreads in the network from peer-to-peer 
according to some contagion rule, 
until reaching a final influence, i.e., a final number of influenced nodes. If the contagion rule is stochastic, then the influence of every seed-set has some distribution, and different seed-sets may have different distributions. 
The algorithmic 
question of finding the seed-set for which the expected influence is maximal has been formalized in \cite{Kempe:2003:MSI:956750.956769} as the \textit{influence maximization problem}, which is NP-hard for a wide family of influence spreading models, and a large body of work has since discussed the development of approximation algorithms for its solution, e.g., \cite{Chen2009,Wang2012,Singer:2012:WFI:2124295.2124381,Borgs:2014:MSI:2634074.2634144,PhysRevX.4.021024,AAAI113670,Wang:2010:CGA:1835804.1835935,jung2012irie,lei2015online,kimura2006tractable}.

While of course all the NP hardness results concern the worst case scenario, we are interested in studying the behavior in the ``typical'' case: 
Is this problem actually hard? 
How do the parameters of the system affect the empirical hardness? %
How much worse will we do if we simply take a random choice? And when does one gain from investing in optimization?

Based on the answers we find to these questions, we also look at their implications from a utility-maximization point of view. 
Each influenced node gives the optimizer some value, the optimization procedure has some cost, and the total utility needs to be maximized. 
It turns out that in many cases optimization is similar to trying to find ``a needle in a stack of needles:'' the choice is among similar outcomes, and the cost of optimization exceeds its marginal profit.

\vspace{5pt}
\noindent {\bf Evaluation metrics and models:  }
We compare the empirical performance of the greedy \textit{hill-climbing} algorithm \cite{Kempe:2003:MSI:956750.956769}, which is known to have the best performance guarantee, 
to a simple randomized choice of a seed set. 
The empirical performance of a seed-set is taken as the median influence gained by this set. 
Denote the empirical performance of the optimized set (using hill-climbing) as \textit{opt}, and the empirical performance of a random choice as \textit{rand}. 
A main metric that we consider is the \textit{marginal gain} from optimization, defined as the difference: $opt-rand$, which is the typical number of influenced nodes added by the optimization process above the random choice benchmark.

Before going into our results, let us describe shortly the models and data that we use. 
The spreading model we consider is the standard independent cascade model, as described in \cite{Kempe:2003:MSI:956750.956769}, where the interaction between nodes is characterized by a parameter $p$, which is the probability of an influenced node to affect each of its neighbors (see Section \ref{influence_model} for more details). Under this model, we evaluate the marginal gain over both synthetic and real-world networks using direct computational experiments. 
%
%
For each network, either synthetic or empirical, we study the influence behavior as a function of the contagion probability $p$. 
We present results over three families of synthetic network models: 
Erd\H{o}s - R\'{e}nyi networks (ER), small world networks (SW) and configuration-model networks with power-law degree distributions (CM). The datasets for these models include networks with 1,000 to 10,000 nodes, and for each network size the results were tested over an ensemble of networks of same type.\footnote{The differences between individual networks of each size were minor. See Section \ref{sec:random} for more details.}
%
We then present results over an empirical dataset of five real-world collaboration networks of different sizes, representing paper co-authorship relations in different arXiv categories. One of these networks (the High Energy Physics Theory network) was discussed also in \cite{Kempe:2003:MSI:956750.956769} in the context of influence maximization. 
See Section \ref{sec:When_Do_We_Need} and \cite{leskovec2007graph,snapnets} for a fuller description of these networks. 

Statistical mechanics considerations 
show that each of the synthetic networks has a percolation phase transition, meaning that there is a {\em critical point} in the parameter space, $p_c$, at which the influence shifts from local to network scale influence \cite{newman2001random,molloy1998size}. 
For empirical networks of course there are no general theoretical predictions, 
but in practice, all the networks in our empirical dataset have an approximate phase transition. 
%
%
%
The results we present are closely related to the phase transition and are similar across all network types, despite their very different degree distributions and structural properties.  
For clarity of presentation, in the introduction we show the results for ER networks and the High Energy Physics Theory collaboration network, and refer the reader to 
Section \ref{sec:When_Do_We_Need} 
for the results of the other network models and empirical networks, and for further details about the datasets and experiments.

\begin{figure}[t!]
\begin{subfigure}{.49\linewidth}
  \includegraphics[width=1.00\linewidth]{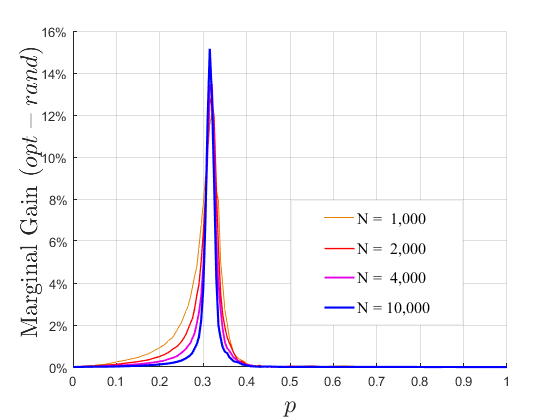}
  \caption{Erd\H{o}s - R\'{e}nyi Networks \vspace{1pt}}  
  \label{fig:median_marginal_gain}
\end{subfigure}
\begin{subfigure}{.49\linewidth} 
  \includegraphics[width=1.00\linewidth]{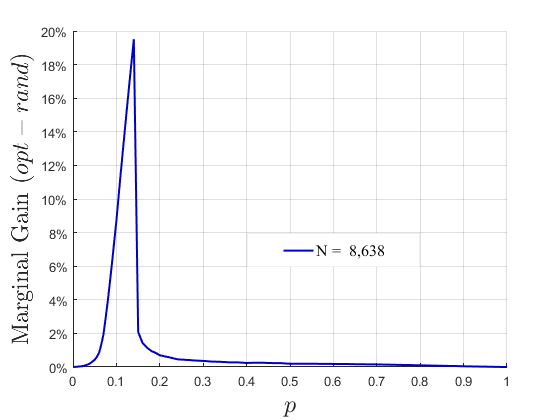}
  \caption{Empirical: High Energy Physics Theory Network}
  \label{fig:median_marginal_gain_empirical}
\end{subfigure}%
\caption{Marginal gain from optimization in respect to the random benchmark, presented as percentages of the network size $N$ as a function of the contagion probability $p$. \textbf{(a)} For ER networks of different sizes and a mean degree of 3. \textbf{(b)} For the collaboration network of High Energy Physics Theory papers on arXiv \cite{leskovec2007graph,snapnets}. Similar results occur for all other synthetic and empirical networks (see Section \ref{sec:When_Do_We_Need}).}
\label{fig:narrow_localization_both}
\end{figure}

%

\begin{figure}[t!]
\begin{subfigure}{.49\linewidth}
  \includegraphics[width=1.10\linewidth]{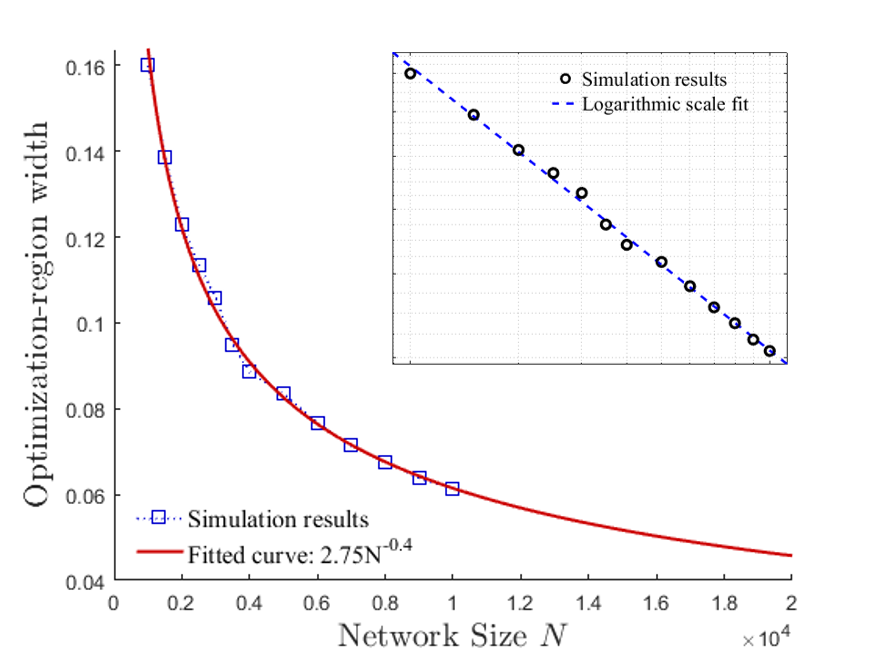}
  \caption{Erd\H{o}s - R\'{e}nyi Networks}  
  \label{fig:optimization_width}
\end{subfigure}
\begin{subfigure}{.49\linewidth} 
  \includegraphics[width=1.10\linewidth]{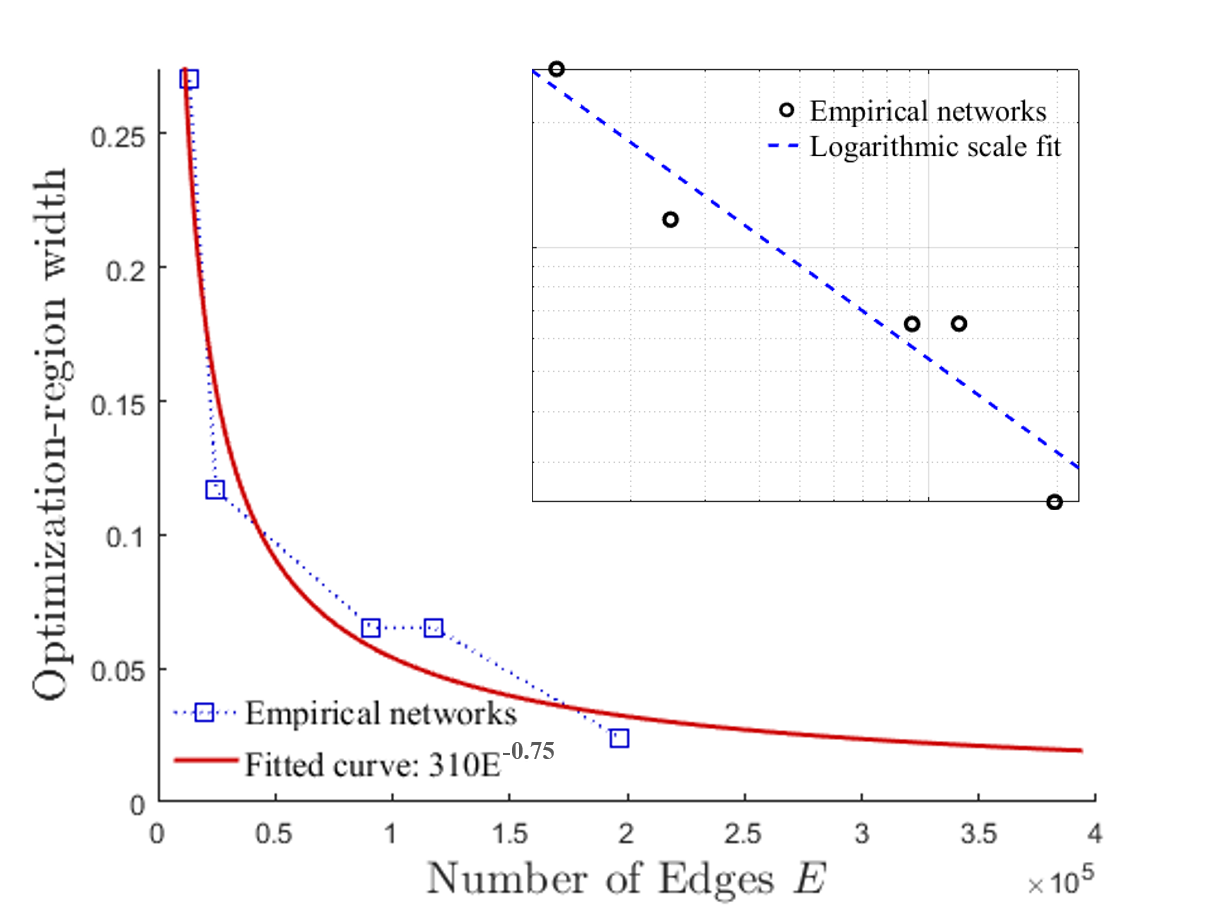}
  \caption{Empirical Networks}
  \label{fig:optimization_width_empirical}
\end{subfigure}%
\caption{Width of the parameter region where the marginal gain from optimization is at least 1\% of the nodes, as a function of the network size. The data points are simulation results and the line is the fitted power-law curve. \textbf{(a)} Optimization-region width in ER networks of sizes $1,000$ to $10$,$000$ as a function of $N$, with  a fitted curve: $A\cdot N^{-a}$, $A = 2.75\pm 0.4$, $a = 0.4 \pm 0.02$. \textbf{(b)} Optimization-region width in empirical networks as a function of the number of edges $E$, with a fitted power-law curve: $B\cdot E^{-b}$, $B = 310\pm 50$, $b = 0.75 \pm 0.3$. The insets show the logarithmic scale fit for each dataset.}
\label{fig:w_vs_size_both}
\end{figure}


\vspace{5pt}
\noindent {\bf Better than random -- only in a vanishing parameter range:} 
We start by characterizing the cases in which influence maximization requires optimization, i.e., 
 where optimization gives significantly better results than a random choice. 
We show 
that 
these cases are centered at a narrow region of the parameter space, near the critical point. 
Figure \ref{fig:median_marginal_gain} shows the 
localization near the critical point of the 
marginal gain for ER networks of different sizes. 
As can be seen, the parameter-space region where the marginal gain is substantial becomes narrow for the larger networks.
Figure \ref{fig:median_marginal_gain_empirical} shows 
the results on the High Energy Physics Theory collaboration network. 
Here again the curve is narrowly localized at the critical region, with only $11$ percent of the parameter space yielding above one percent marginal gain, and a less than $4$ percent of the parameter space yielding above $10$ percent marginal gain. 
The results for the other synthetic and empirical networks were similar (but with different critical points for each network, as theoretically expected for networks with different degree distributions), and again the width becomes narrow for the larger networks (see Section \ref{sec:random} for the results on the full empirical network dataset and Section \ref{sec:SW_and_SF}
 for the results on SW and CM networks).

A natural way to measure the width of the parameter region where optimization is needed, is to test for which values of $p$ the marginal gain from optimization is 
at least $1\%$ of the network size $N$. 
We show that 
the width of this region vanishes as a power law 
of the network size. 
Figure \ref{fig:optimization_width} shows that for the ER network dataset, the width decreases as $N^{-a}$, 
with $a\approx0.4$.   
In Section \ref{sec:SW_and_SF} we show that similar scaling behavior holds for the other synthetic network models. 
Figure \ref{fig:optimization_width_empirical} reveals a similar picture for the empirical networks, again exhibiting a power-law vanishing of the critical region where optimization may be significant. Since these networks have different edge densities, the relevant size yielding a power-law shrinking in this case is the number of edges $E$, rather than the number of nodes. We find that the width in the empirical dataset scales as $E^{-b}$, with $b\approx0.75$. See Section \ref{sec:random} for more details.  

\begin{figure}[t!]
\begin{subfigure}{.49\linewidth}
  \includegraphics[width=1.10\linewidth]{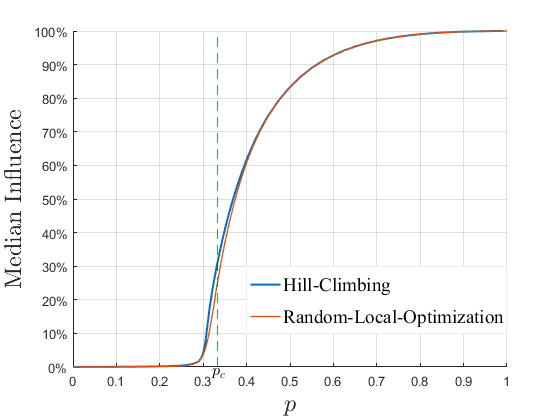}
  \caption{Erd\H{o}s - R\'{e}nyi}  
  \label{fig:HC_and_random_local}
\end{subfigure}
\begin{subfigure}{.49\linewidth} 
  \includegraphics[width=1.10\linewidth]{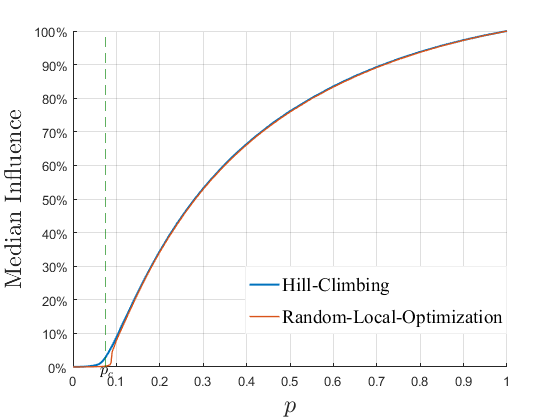}
  \caption{Empirical}
  \label{fig:HC_and_random_local_empirical}
\end{subfigure}%
\caption{Median influence of the seed-sets obtained by hill-climbing and random-local-optimization as percentages of the network size. \textbf{(a)} For Erd\H{o}s - R\'{e}nyi network of size $N=$ 10,000 and a mean degree of 3. \textbf{(b)} For the High Energy Physics Theory collaboration network. The advantage of hill-climbing is centered near the phase transition at $p_c$.}
\label{fig:HC_and_random_local_both}
\end{figure}

 
The practical meaning of these results, and their empirical robustness across network types, is that for large scale networks, optimization is likely to give substantial gains only if the system is near its critical point, in a region that approaches a zero measure of the parameter space. To the best of our knowledge, we are the first to 
study the effects of optimization near the critical point and to 
identify 
the phase transition as the underlying reason for the differences between different seed-sets, and for any significant sub-optimality of a random choice -- and thus the need for optimization. 
 
\vspace{5pt}
\noindent {\bf Economic implications:  }
Next, we turn to look at the gained \textit{utility}, i.e., 
considering also the optimization costs, where the cost is proportional to the optimization running time. 
We consider the case where both the gains for the optimizer from influenced nodes and the costs incurred by the optimization process can be quantified on the basis of the same monetary units. 
We find that away from the critical point, the marginal gain from running optimization algorithms in respect to the random benchmark is small, and does not increase with the network size
, while the cost of optimization does increase with the network size; as a result, for every given cost of computation-per-time-unit, in large enough networks the price of optimization exceeds its added profit and the added utility from optimization in respect to the random benchmark becomes negative. This leaves only a narrow parameter region near the phase transition in which the marginal gain does scale with the network size, and ``it pays'' to run optimization procedures. 
Specifically, considering the nearly-linear running time obtained by state-of-the-art algorithms (e.g., \cite{Borgs:2014:MSI:2634074.2634144,Wang2012}), the width of the region where optimization gives a positive added utility diminishes as a power law of $N$. I.e., for almost all of the parameter space optimization procedures on large networks result in negative added utility. See Section \ref{subsection:Considering optimization costs} for the full analysis and results.   

\vspace{5pt}
\noindent {\bf Constant-time optimization:  }
The results described above indicate that 
for large-scale networks global optimization procedures are wasteful in almost all of the parameter space, 
and even linear time algorithms may become too expensive. 
To address this problem, 
we introduce a constant-time optimization approach that achieves high 
influence performance,  
while maintaining constant computation 
costs which are independent of the network size. 

We demonstrate this approach through a simple algorithm we call \textit{random-local-optimization}.  
The basic idea is that in order to find a seed set of $k$ nodes, the algorithm constructs sub-networks of constant size $M$ around $k$ randomly chosen nodes, and then optimizes over each sub-network separately. 
Figure \ref{fig:HC_and_random_local_both} shows the median influence achieved by the random-local-optimization and by the hill-climbing algorithms, as a function of the parameter $p$, for ER networks of size $N=$ 10,000 and for the empirical High Energy Physics Theory collaboration network. 
We see that in terms of the influenced-set size, the simple and fast random-local-optimization algorithm achieves similar performance to hill-climbing in most of the parameter space, except near the critical point, where hill-climbing has a modest advantage. This result was robust across all synthetic and empirical datasets.
Moreover, in Section \ref{section:Constant time optimization} we show that in terms of the gained utility, the random-local-optimization algorithm outperforms the standard optimization results for the case of large networks and unknown parameters, even when considering the running-time costs of state-of-the-art optimization methods.  
See Section \ref{section:Constant time optimization} for more details of our algorithm and the evaluation results. 

This constant time approach may be applicable to large networks where global optimization is not practical, or to natural partial-information scenarios where the system parameters are unknown. The short computation time obtained this way is also important for time-varying networks, in which global computation which leads to long solution times may render the solutions irrelevant.

\section{Preliminaries} \label{sec:preliminaries}
\subsection{The Influence Maximization Problem}
The influence maximization problem was first discussed in the context of marketing \cite{Domingos:2001:MNV:502512.502525}, and later formalized in the seminal work of Kempe, Kleinberg and Tardos (2003) as a general algorithmic problem, with many papers to follow, including works from algorithmic game theory and mechanism design perspectives \cite{seeman2013adaptive,Singer:2012:WFI:2124295.2124381}, targeted immunization \cite{PhysRevX.4.021024,giakkoupis2005models}, competing influences \cite{dubey2006competing,carnes2007maximizing,bharathi2007competitive}, and more (for reviews, see \cite{Lue2016,v011a004}). The formal algorithmic problem is simple: \textit{for a parameter $k$, find the $k$-node seed-set that gives the maximal average influence}. Simple as it may seem, the influence maximization problem is NP-hard for a family of cascade and threshold influence models, and the vast majority of the related literature discusses approximation algorithms for its solution.
The closest related work 
is \cite{akbarpourjust} that studies the possibility of using randomized seeding; however, the study focuses on 
the different question of   
the effect of using different sizes of target sets, and the analysis does not concern optimization near the phase transition 
and does not apply to finite systems. 
%
%
The present study focuses on the effect of the phase transition and its finite-size critical region on the optimization problem, and on the consequent scaling with the network size of the parameter region where optimization is indeed practically hard. 
%
%

The practical hardness of the problem is different from its formal one. Since there are only ${N}\choose{k}$ $=\mathcal{O}(N^k)$ sets of size $k$, for a \textit{fixed} constant $k=K$, the task of finding the maximal-influence set is polynomial in the network size $N$,  
and although finding the exact average influence of a seed-set is known to be \#P-hard \cite{Wang2012}, approximating it well is easy by a direct Monte-Carlo simulation (see the proof regarding sampling time in \cite{v011a004}). However, for real-world purposes where networks may consist of billions of nodes, exhaustive search or sampling solutions are not practical even for small $k$ 
in spite of being polynomial. The practical hardness of the problem stems from the sheer size of the networks one may need to work with, and even linear-time algorithms may become expensive -- depending on the actual networks in discussion. This provides motivation to seek sub-linear algorithms for this problem, as we suggest in Section \ref{section:Constant time optimization}. 

\subsection{The Influence Model and the Phase Transition}\label{influence_model}
We consider the independent cascade model of network spreading, as described in \cite{Kempe:2003:MSI:956750.956769} (see also \cite{SOLOMON2000239,Goldenberg2001}).  
The idea of the model is that each link between two nodes is characterized by some probability, and each influenced node tries only once to affect each of its neighbors, and succeeds with the probability associated with the link between them. We will consider a uniform probability $p$ for all links. The process progresses in discrete time steps, and after an influenced node tries to affect all its neighbors, it becomes inactive in the following time steps (but is still counted as an influenced node). The model is equivalent to the discrete SIR model \cite{kermack1927contribution}, studied in the context of epidemic spreading, with a removal probability of $1$. 
We also tested a variant of the model in which the contagion probability $p$ is not uniform, but instead the contagion probabilities through each of the edges, $\left\{p_{ij}\right\}$, are randomly distributed with a Gaussian distribution centered at a value $p$, which is a property of the network and of the message being spread. 

A useful way to look at the diffusion of influence in the network, is the \textit{static picture}, as opposed to the dynamic picture described above. In the static picture, instead of tossing a coin at each step of the process to see whether a link is conductive or blocked, all coins are tossed at the beginning, and the network is broken into clusters. The two pictures are equivalent, and this equivalence has been noted and used for proving the sub-modularity of the influence function in \cite{Kempe:2003:MSI:956750.956769}. In addition, the static picture also points out to the similarity of our problem to the well studied field of bond percolation (see, e.g., \cite{aharony2003introduction,grimmett1999percolation}). 
Relevant results from percolation theory are the existence of a critical threshold $p_c$, at which a giant cluster appears (i.e., a cluster of the order of the network diameter), with a power-law distribution of the cluster sizes near this critical point, and the scaling of the average cluster size, which scales as $\left|p-p_c\right|^{-\gamma}$ near the critical point, with $\gamma > 0$. 
The influence function discussed in the influence maximization problem is the average over the cluster size distribution around each specific set of nodes. The questions of whether, and to what extent, is there a difference between various initial seed-sets, depend on the extent to which the cluster sizes are correlated to specific network sites, and on the cluster size distribution. The distribution depends on the parameter $p$, which plays a central role in determining who are the important nodes, and how much are they different than the average nodes.       

\subsection{Examples: The Importance of the Influence Parameters}
Let us start with two motivating examples that demonstrate the importance of the influence parameters for the result of the optimization process, and show the interplay between these parameters and the network structure.
We look at small toy networks where the influence can be easily solved exactly, but their structure still demonstrates basic phenomena that play an important role in influence maximization in large and complex networks as well.

\begin{figure}[t!]
\centerline{
\begin{tabular}{cc}
\includegraphics[width=1.05\linewidth]{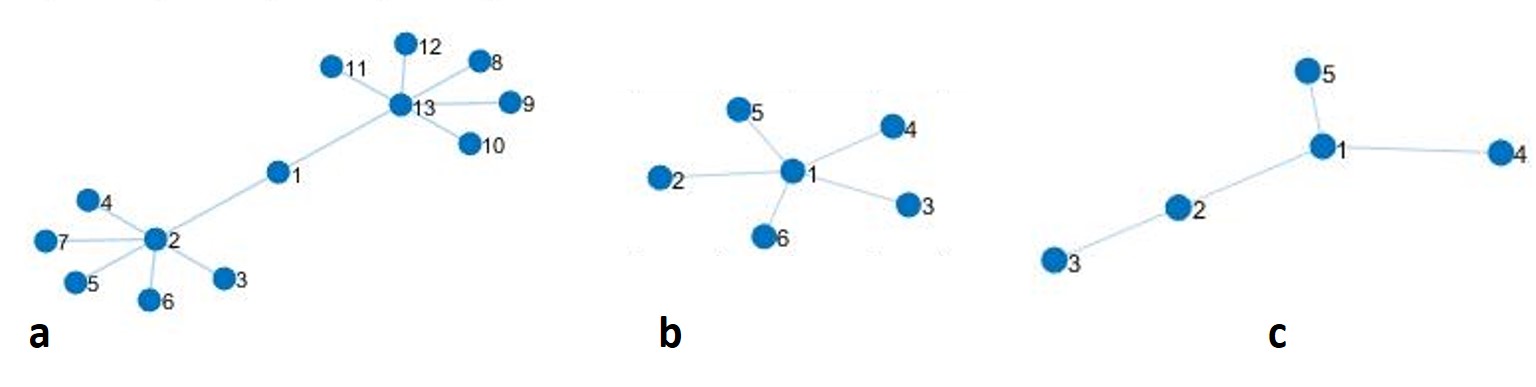}
\end{tabular}
}
\caption{Network \textbf{(a)}: The maximal-influence node depends on the parameters; the best node is node 1 for $p>4/5$, but it is node 2 or 13 for $p<4/5$. Network \textbf{(b)}: The central node has the highest influence in every realization. Network \textbf{(c)}: The best node is not the best in every realization: Node 1 has the maximal influence in this network only in a $\left[1-p(1-p)^3\right]$ fraction of realizations.}
	\label{fig:toy_networks}
\end{figure}
%
Our first example demonstrates that the best set of a given network depends on the influence parameters. 
Assume we try to find a single seed node in the network in Figure \ref{fig:toy_networks}(a). Every link has a  probability $p$ to be conductive, and the average influence of every node can be calculated.\footnote{This kind of calculation is manageable in tree networks; if there are cycles such calculation will demand counting all non-backtracking trajectories between all seeds and targets, which becomes unfeasible in large networks. } The result is that if $p>4/5$, then the best node is node $1$, who has only two neighbors, while for $p<4/5$, the optimal nodes are nodes $2$ or $13$, who are equivalent with $6$ neighbors each. 
This illustrates how in real networks, large hubs can be important when the influence probability is below some threshold, but if the probability is higher, low degree nodes who are close to several large hubs may become more important than the hubs themselves. Therefore, the question of ``who are the most influential nodes?'' should in fact be answered by: ``What are the influence parameters?''. 

Our second example demonstrates that the desired best seed-set is not always the best one in each specific scenario, i.e., it does not  give the maximal influence in every realization of the random process.
The star network in Figure \ref{fig:toy_networks}(b) is a non-typical case where the most influential node does have the maximal influence in every realization. In contrast, in the network in Figure \ref{fig:toy_networks}(c), node $1$ has the best average influence, but there is a probability of $p(1-p)^3$ in each realization that nodes $2$ and $3$ will have higher influence. In more complex networks the differences between many of the seed-sets may be small, and this gives hope that some partial and fast optimization may suffice to yield good results, as there may be many ``nearly best'' seed-sets in the network.

\section{When Do We Need Optimization?}\label{sec:When_Do_We_Need}
Our goal in this section is to characterize the cases in which optimization procedures are required. 
The answer depends both on the networks in discussion and on the influence parameters. 
Special networks can be constructed such that the optimization can be dramatically important, as in the case of a star network, or completely useless, as in the case of a uniform lattice where all nodes are equivalent, and the solution is every set with uniform distances between its nodes.  
However, such special cases are far from being typical representations of real-world networks, and in order to get a more general picture we look at different types of random networks and at empirical networks, and focus our question on the effect of the system's parameters on the necessity of optimization.

As explained in the previous section, the strength of peer-to-peer interaction is characterized by the parameter $p$, which is the probability of an influenced node to affect each of its neighbors. We  measure the effects of optimization for different values of this parameter, and ask: For which range of parameters ``it pays'' to run optimization algorithms?  
\subsection{Doing Better Than Random} \label{sec:random}
A natural way to test whether and when do we need optimization algorithms is to compare to the random choice benchmark. Optimization takes time and effort and it will be useless if it gives similar results to an uninformed shot in the dark. We test for which parameters does optimization using the hill-climbing algorithm proposed in \cite{Kempe:2003:MSI:956750.956769} give higher influence than a random choice, and measure the marginal gain from optimization. 
%
We denote the median influence 
%
%
gained using the hill-climbing optimization procedure\footnote{For most of the parameter range the median and the mean estimates coincide. However, near the phase transition the cluster sizes have a power law distribution, and the median as a robust estimator gives a more reliable estimate for the typical influence.} 
as $opt$, and the median influence of the random choice as $rand$, and evaluate the marginal gain, $opt-rand$, as a function of the parameter $p$.

The data for the experiments consist of three datasets of synthetic networks and a dataset of five real-world networks. 
Each synthetic network dataset includes networks of $10$ sizes, ranging from 1,000 to 10,000 nodes, and 10 network instances of each size. For each of the network types independent experiments were conducted over a grid of 85 unequally spaced $p$ values, with intervals of $0.0025$ near the phase transition where a higher resolution is required, and up to $0.025$ at the low and high phases where the influence derivative as a function of $p$ is small. For each network instance of each size and for each $p$ value we ran 20,000 independent experiments of the influence spreading process, which according to preliminary measurements provided a sufficient sample for all networks. The real-world network dataset includes $5$ networks of different sizes, ranging between 4,158 and 21,363 nodes and between 13,428 and 197,031 edges; the experiments on these networks were conducted with similar grids of $p$ values and number of experiments as for the synthetic networks. 

\begin{figure}[t!]
\begin{subfigure}{.49\linewidth}
  \includegraphics[width=1.10\linewidth]{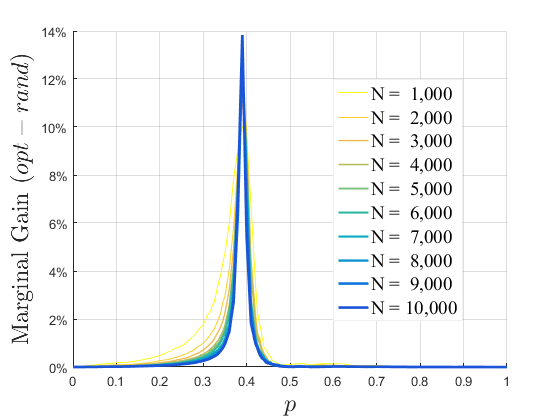}
  \caption{ }  
  \label{fig:SW_narrow_localization}
\end{subfigure}
\begin{subfigure}{.49\linewidth} 
  \includegraphics[width=1.10\linewidth]{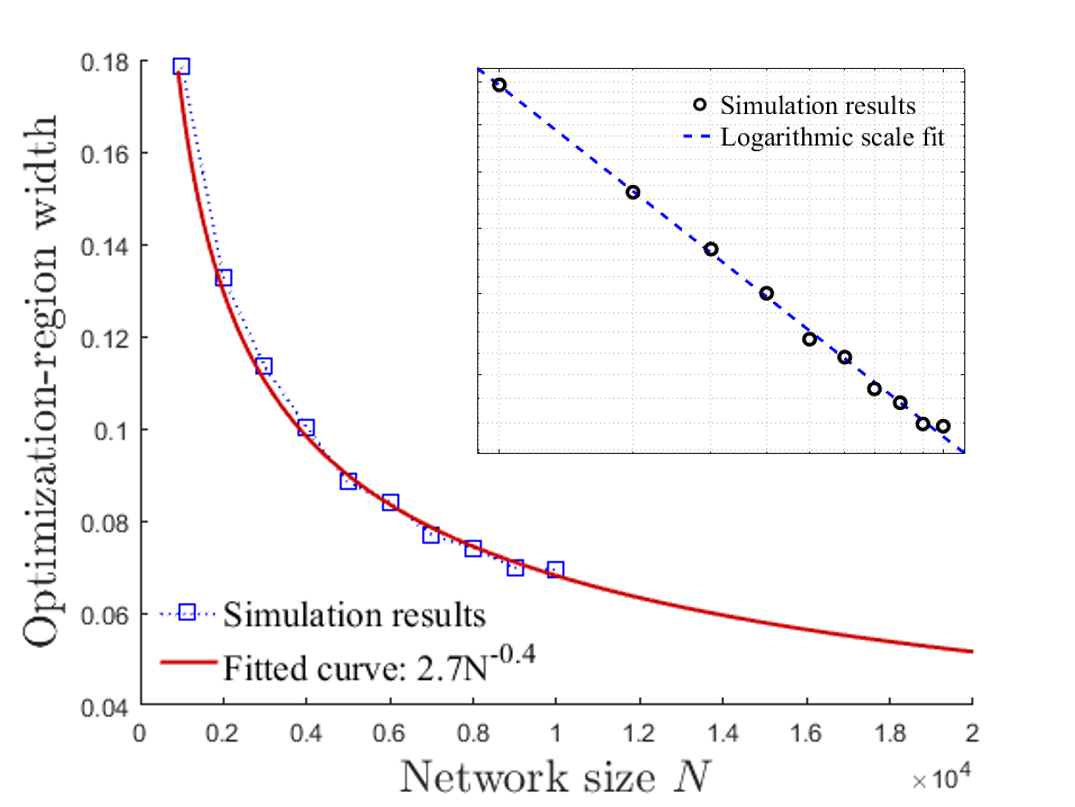}
  \caption{ }
  \label{fig:SW_optimization_width}
\end{subfigure}%
\caption{Small-world networks.  
	The networks were generated according to the Watts and Strogatz model with a rewiring probability of 0.2 and a mean degree of 4.  \textbf{(a)} Marginal gain from optimization in respect to the random benchmark, presented as percentages of the network size. 
	\textbf{(b)} Width of the parameter region where the marginal gain from optimization is at least 1\% of the nodes, as a function of $N$. The data points are simulation results and the line is the fitted power-law curve: $A\cdot N^{-a}$, $A = 2.7\pm 0.7$, $a = 0.4 \pm 0.03$. The inset shows the linear fit in logarithmic scale.}
\label{fig:small_world_figures}
\end{figure}

\begin{figure}[t!]
\begin{subfigure}{.49\linewidth}
  \includegraphics[width=1.10\linewidth]{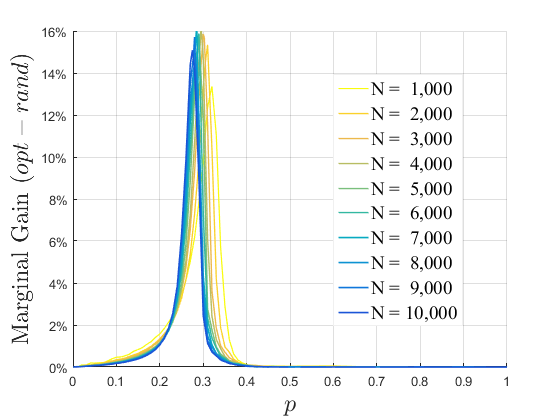}
  \caption{ }  
  \label{fig:SF_narrow_localization}
\end{subfigure}
\begin{subfigure}{.49\linewidth} 
  \includegraphics[width=1.10\linewidth]{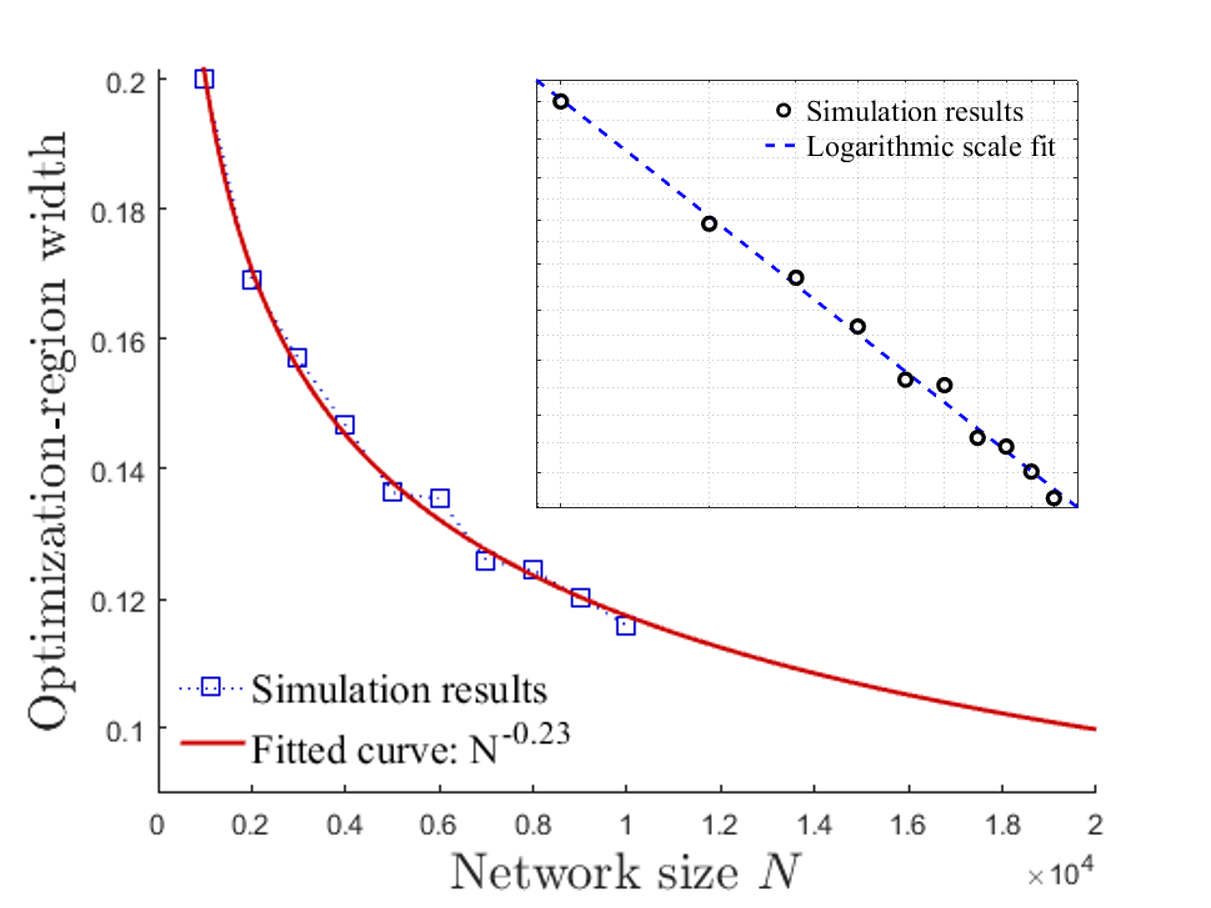}
  \caption{ }
  \label{fig:SF_optimization_width}
\end{subfigure}%
\caption{Power-law configuration model networks. 
	The networks were generated with a power-law degree distribution with an exponent of 2.5 and a lower cutoff of 4. 
	\textbf{(a)} Marginal gain from optimization in respect to the random benchmark, presented as percentages of the network size.
	\textbf{(b)} Width of the parameter region where the marginal gain from optimization is at least 1\% of the nodes, as a function of $N$. The data points are simulation results and the line is the fitted curve: $A\cdot N^{-a}$, $A = 1\pm 0.4$, $a = 0.23 \pm 0.03$. The inset shows the linear fit in logarithmic scale.}
\label{fig:scale_free_figures}
\end{figure}

Figure \ref{fig:median_marginal_gain} shows the marginal gain on Erd\H{o}s - R\'{e}nyi networks of different sizes, with a mean degree of $3$ and an initial seed-set of size $k=3$. As can be seen, the marginal gain is narrowly localized  in a limited region of the parameter space. 
%
%
The narrow localization of the marginal gain was persistent for other choices of mean degree, and was similar across other random network models, as we discuss in section \ref{sec:SW_and_SF} (see Figures \ref{fig:small_world_figures} and \ref{fig:scale_free_figures}). The difference is in the location of the peak which is near the critical point of each network. All the results were robust also to different seed-set sizes; in Appendix \ref{subsection:k_vals} we discuss the effect of different $k$ values, and show that this effect conforms with theoretical predictions derived from percolation theory. 
It is important to note in this regard that 
near the phase transition, in the parameter range predicted by our power-law scaling results, 
for any value of $k$ 
optimization is still better than a random choice (as long as $k$ itself does not cover a significant portion of the network). This is since near $p_c$ the number of large clusters diverges with $N$, and results that rely on the existence of a single giant component, or the existence of only logarithmic clusters, are no longer valid. 
%
%

 \begin{figure}[t!]  
\centerline{
\includegraphics[width=1.05\linewidth]{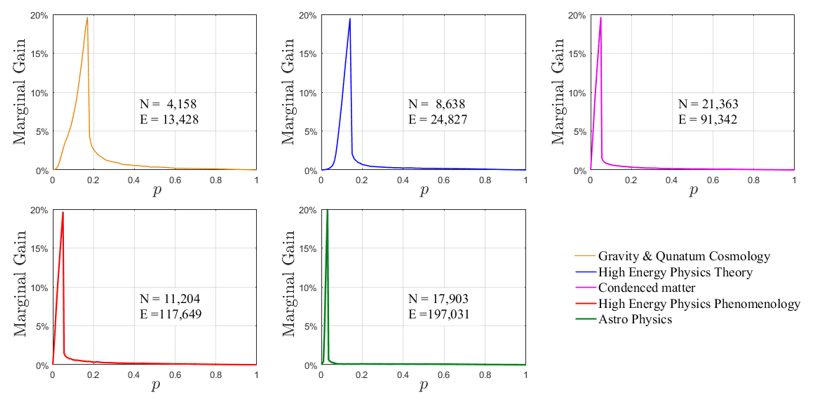}
}
\vspace{-10pt}
\caption{Marginal gain from optimization in respect to the random benchmark in percentages of the network size $N$, for the dominant component of each of five empirical collaboration networks of papers on arXiv from different categories \cite{leskovec2007graph,snapnets}. The legend states the arXiv category of each network, and refers to the plots from left to right and top to bottom.}
	\label{fig:SW_narrow_localization_five_networks}
\end{figure}     

In the empirical networks as well, we observe a narrow localization of the marginal gain curve in a single peak near the critical point of each network. 
Figure \ref{fig:SW_narrow_localization_five_networks} shows the marginal gain curves as a function of $p$ for the dominant components of the five empirical collaboration networks. In these networks the nodes represent authors of papers on arXiv in five different categories, and an edge exists between two authors if they have co-authored at least one paper in that category. The numbers of nodes and edges in each network are denoted in the figure as $N$ and $E$, respectively. For a detailed description of these networks and their properties see \cite{leskovec2007graph,snapnets}. 
The critical points of empirical networks generally do not have known theoretical values, however they can be evaluated using standard methods.\footnote{E.g., by using a parametric fit of the degree distribution of each network to an approximate function, and applying the results of \cite{newman2001random,callaway2000network} to calculate the critical point, or by a direct measurement of the size of the largest cluster at the high phase in each realization as a function of $p$ and an extrapolation of the average maximal size.} 

The practical meaning of the results presented in Figures \ref{fig:narrow_localization_both} and \ref{fig:SW_narrow_localization_five_networks} is that optimization is substantially better than a random choice only in a specific region of the parameter space, near the critical point; 
in all networks, the marginal gain has a peak centered near the critical point, as calculated from percolation theory. 
The marginal gain is close to zero both above and below this peak, which can be intuitively understood in the limits of low and high probabilities: at low contagion probabilities, spreading is local and is related mainly to first neighbor influence. In this regime only small clusters exist in each realization, and the possibilities for optimization are limited. At high probabilities, at the end of each realization of the random process almost all nodes are connected as one giant influenced cluster, while all other clusters are again small; the influence is then high, however a random choice also has a high probability to hit this same cluster and gain the same influence -- and so again optimization is not much better than a random choice. Only near the phase transition there is a power-law distribution of the cluster sizes and hence a large variance, which allows for significant differences in influence between network sites. The narrow localization of the curve in large networks is also consistent with finite-size scaling results from statistical mechanics, showing that the transition region below $p_c$ where the power-law scaling of the clusters is valid, diminishes as a power law of $N$ \cite{finite-size-scaling-in-percolation,9de370b853f048c8ac85a09335c1d767}.

%

In order to measure the width of the parameter range where optimization is needed (the ``optimization region''), we test for which values of $p$ the marginal gain from optimization is at least $1\%$ of the nodes. 
We find a clear scaling behavior, and show that the width of this region decays to zero as a power law of the network size. 
Figure \ref{fig:optimization_width} shows the optimization-region width as a function of the network size for the ER networks, and 
the inset shows a power-law fit in logarithmic scale with an exponent of $0.4 \pm 0.02$. Each data point is the width in parameter space in which the marginal gain is at least 1\% of the network, averaged over 10 networks of each size. The differences between the width results calculated on each network separately for each size were small, and were mainly near the phase transition, with maximal standard deviations of $0.009$ to $0.015$ 
for $N=$ 10,000 to 1,000, respectively. For most of the parameter range (away from $p_c$) the differences between networks were an order of magnitude smaller.   
We also compared to other methods to estimate the width of the marginal gain curve -- the standard deviation and the full-width-at-half-maximum of the marginal-gain curves -- and both gave the same results for the exponents. Figure \ref{fig:optimization_width_empirical} shows the results for the empirical collaboration networks. Here, since these networks have different edge densities, the relevant size is the number of edges $E$. In this empirical dataset as well, we see a power-law shrinking, with an exponent of $0.75 \pm 0.3$.
The noisy variant of the contagion model that we have tested, where the contagion probabilities are no longer uniform but normally distributed around an average value $p$ with a variance of $0.1$, did not change any of results described above, and specifically,  the power-law scaling exponents remained the same. 


%
%

\subsection{Power-Law Configuration-Model Networks and Small-World Networks}\label{sec:SW_and_SF}
We have seen in 
the previous section that for both ER networks and real-world networks the marginal gain curves are narrowly peaked near the critical points, and that the optimization region vanishes as a power law of the network size. 
Here, we show that these results hold also for other network types. This is despite significant differences between the networks (in terms of degree distributions, diameters, local symmetry, etc.), and the different processes in which they were generated.
Next, we shortly describe the power-law configuration-model and small-world synthetic network models, and show the results over these datasets.

\vspace{5pt}
\noindent {\bf Small-world networks: } The small-world networks that we use were generated according to the Watts and Strogatz model \cite{watts1998collective}. In this model, the network is initialized as a regular lattice with periodic boundary conditions and $N$ nodes, where each node has $z$ neighbors (often called the coordination number). Then, each edge in the network has a probability $\mu$ to be rewired to a random target node under the constraint that there are no self loops and no duplicate edges. Networks generated in this process interpolate between a regular graph ($\mu=0$) and a random graph ($\mu=1$). For intermediate values of $\mu$ the networks have the property of a small diameter, as in random graphs, but with a higher clustering coefficient and a highly symmetric local neighborhood structure.
See \cite{watts1998collective} for more details about the model.

The results for these networks are presented in Figure \ref{fig:small_world_figures} for networks with $z=4$ and $\mu=0.2$.
These results were robust for different choices of mean degrees and rewiring probabilities, only with different values of the critical point. Figure \ref{fig:SW_narrow_localization} shows the marginal gain as a function of $p$ in percentages of the network size, for networks of sizes 1,000 to 10,000, and Figure \ref{fig:SW_optimization_width} shows the power-law shrinking of the optimization region with the network size. Specifically, the width of the optimization region is proportional to $N^{-a}$ with $a = 0.4 \pm 0.03$. 
We note that this is the same exponent obtained for the ER networks. From a statistical mechanics point of view, this result is predicted, since for every finite $\mu$ the SW and ER networks are known to be of the same universality class \cite{barrat2000properties}, meaning that they have similar scaling behavior near the phase transition.


\vspace{5pt}
\noindent {\bf Power-law networks: }
We generated the dataset of (finite) power-law networks using a randomized configuration model \cite{molloy1995critical, molloy1998size, newman2001random}. In this model, a sequence of degrees $(k_1,...,k_N)$ for the $N$ nodes is drawn independently from a given probability distribution\footnote{Note that $\sum_i k_i$ must be even. This could always be fixed by adding one edge to a random node.} $Pr(k)$, and then a graph is drawn uniformly at random from the set of all graphs with that specific degree sequence. 
A convenient implementation of this model is to consider each node $i$ to have exactly $k_i$ half edges; then, pairs of half edges are randomly connected under the constraint of no self loops and no duplicate edges. This implementation is not guaranteed to be solvable (i.e., to yield a graph which satisfies the degree sequence and the other constraints), however, if the process reaches an impasse, one may randomly rewire a small number of edges, or alternatively restart the whole process. In our dataset this happened only in a small number of cases, in which we then applied the second approach of a full restart. 

Figure \ref{fig:scale_free_figures} shows the results for configuration model networks of sizes 1,000 to 10,000 and a power-law degree distribution with a power of $2.5$ and a lower cutoff of $4$. In Figure \ref{fig:SF_narrow_localization} we see the marginal gain curve in percentages of the network size for the different values of $p$. The approximate value of the critical point is given by $p_c = \frac{\left\langle k\right\rangle}{\left\langle k^2\right\rangle - \left\langle k\right\rangle}$ \cite{cohen2004fractal}, 
and approaches zero for infinite networks,\footnote{More precisely, $p_c$ approaches zero for networks with a degree distribution $Pr(k)\propto 1/k^\alpha$ with $\alpha < 3$. For power-law networks with $\alpha>3$ the formula is valid but approaches a finite value.} 
and hence the left-shifting of the peaks. Figure \ref{fig:SF_optimization_width} shows the power-law shrinking of the optimization region as $N^{-a}$ with $a = 0.23 \pm 0.03$.


To summarize, despite the differences in local neighborhood structure and the very different degree distributions, the results are similar for all these networks -- in all our synthetic and real-world network datasets, 
optimization is significant only in a vanishing parameter region near the phase transition. 
The main differences between networks are only in the location of the critical point and the value of the shrinking exponent, which could be expected, however the qualitative results, as well as their implications in the large $N$ limit are similar across all network types.

\subsection{Considering Optimization Costs}\label{subsection:Considering optimization costs}
In order to get a more concrete picture of whether and when optimization procedures are profitable, we should also take into account the costs of optimization, and to see when does one have a positive added utility from running optimization procedures. The following results are a corollary of the above results, that optimization has a significant marginal gain only in a vanishing region, together with the fact that computation costs increase with the network size. 

Optimization has direct costs resulting from computation time (see, e.g., \cite{Babaioff:2017, 5054870}), as well as possibly other indirect costs incurred by the time delay. We model these costs as a constant cost-per-time-unit $C$, and the total cost\footnote{We omit here any cost incurred by the first $k$ seed nodes starting the process. The question of optimizing the deployment of a budget between costs of the initial seeds is discussed in \cite{Singer:2012:WFI:2124295.2124381}; here we focus on the costs of the optimization process itself.} of optimization is proportional to the running time: $C\cdot T$.  
The optimization time $T$ scales with the network size in the standard methods, with 
state-of-the-art algorithms achieving impressive developments with nearly linear running times; but still, for large networks the optimization cost $C\cdot T$ may be substantial, and if the marginal gain from optimization is small, this cost may exceed the marginal gains. 

More formally, each influenced node $i$ gives the user some value $v_i$, and we consider the simple case of a uniform value for all nodes $v_i=v$. The condition under which optimization yields a positive added utility is $U_{opt} > U_{rand}$. Substituting $U_{opt} = v\cdot opt - C\cdot T$ and $U_{rand} = v\cdot rand - C$ we have the condition:
\vspace{-2pt}
\begin{equation}
\frac{1}{T-1} \left(opt-rand\right) > \frac{C}{v}
\label{eqn1}
\vspace{-2pt}
\end{equation}

We can see the competition between low computation costs on the one hand, and large datasets which demand long running times on the other hand. More importantly, in light of the sharp localization of the marginal gain term $\left(opt-rand\right)$ in the parameter space discussed above (see Section \ref{sec:When_Do_We_Need}), we expect that eq.\ref{eqn1} will be satisfied only in a narrow parameter region; as we show next, this is indeed the case for standard optimization algorithms. 

In order to estimate the width $w$ of the region in which standard optimization gives positive added utility (i.e., when we now also consider the costs), we use simulation results of hill-climbing \cite{Kempe:2003:MSI:956750.956769}, which is proved to have the optimal approximation guarantee, and calculate the optimization costs incurred by the running time of $N{}log(N)$ achieved  by state-of-the-art algorithms.

\begin{figure}[t!]
\begin{subfigure}{.49\linewidth}
  \includegraphics[width=1.10\linewidth]{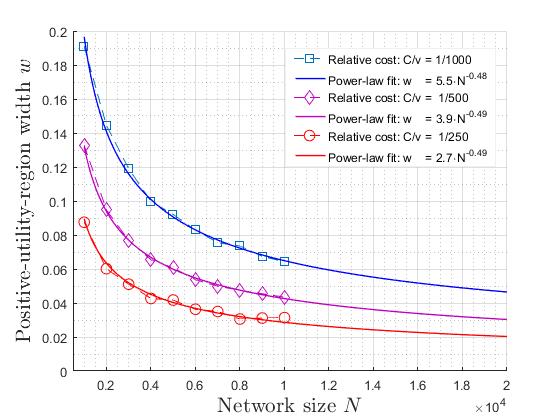}
  \caption{Erd\H{o}s - R\'{e}nyi Networks}  
  \label{fig:w_vs_N_with_costs_ER}
\end{subfigure}
\begin{subfigure}{.49\linewidth} 
  \includegraphics[width=1.10\linewidth]{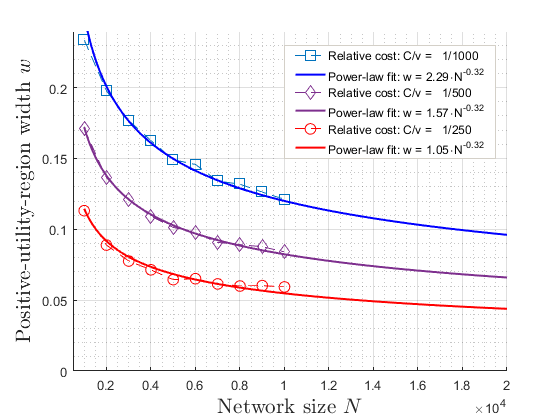}
  \caption{Power-law CM Networks}
  \label{fig:w_vs_N_with_costs_SF}
\end{subfigure}%
\caption{Width $w$ of the region where optimization gives positive added utility above a random choice when considering the hill-climbing algorithm performance and the lower optimization running time of $N{}log(N)$, for different relative costs of optimization: \textbf{(a)} In ER networks; \textbf{(b)} In configuration-model networks with a power-law degree distribution.}
\label{fig:utility_figures_power}
\end{figure}

Substituting this running time and the empirical results for ($opt-rand$) into eq.\ref{eqn1}, we can test for which values of $p$ does the inequality hold; this defines a width $w$ in the parameter space, in which the added utility is positive. We then look at the functional dependance of $w$ on the network size $N$, and extrapolate the results for larger networks. 
Figure \ref{fig:utility_figures_power} shows the results of $w$ as a function of the network size for example values of the relative cost $C/v$. 
Figure \ref{fig:w_vs_N_with_costs_ER} depicts the results for ER networks and Figure \ref{fig:w_vs_N_with_costs_SF} for CM power-law networks. The SW network dataset produced similar results, with the same exponent as the ER networks.
In all cases we find that $w$ has a clear power-law decay with $N$, in consistency with our previous results. Notice that using different values of the relative cost $C/v$ only amounts to a multiplicative factor, and does not change the power-law decrease with $N$.    


%

In the bottom line, the range of parameters in which it pays to use optimization algorithms is narrowing as a power law of the network size. 
If the system happens to be close to its phase transition, 
it is useful to choose the initial seed-set using optimization algorithms; but in  real-world scenarios the parameters may not be known, and cannot be easily tuned, and in these cases, for large networks there is a  small probability to gain anything from running global optimization algorithms.
A practical approach to deal with this unfortunate situation (from an optimization perspective) is to bound the cost of the optimization process such that it will not scale with the network size, and then see how well can we do. We demonstrate this approach in the next section.

\section{Constant Time Optimization}\label{section:Constant time optimization}   
We have seen that standard optimization may be wasteful for most of the possible parameters, yet it may be profitable if the system happens to be close to its critical point.   
Therefore, we suggest an approach that on the one hand, avoids high costs -- for the case that the system is away from its phase transition -- and on the other hand, if the system is close to its critical point, will capture a significant part of the optimization gains, and will 
do better than a random choice, and not much worse than standard optimization. 
The key idea is to maintain constant optimization costs, that are independent of the network size. We demonstrate this approach by a simple algorithm we call \textit{random-local-optimization}.

\vspace{5pt}
\noindent \textbf{Random-local-optimization:} In order to maintain a constant running time it is needed to avoid iterating through the whole network data. The basic idea is that in order to find a seed-set of $k$ nodes, the algorithm chooses $k$ random nodes and builds bounded sized sub-networks around them. The optimization is then performed only over these sub-networks, and for each sub-network independently. Specifically, a user chooses some constant mass $M$, which is the number of nodes in each sub-network, and the algorithm gets this constant as input. The algorithm then takes $k$ random nodes to be the initial $k$ sub-networks. Then, while each sub-network mass is less than $M$, the algorithm incrementally adds the neighbors of each current sub-network. When this process terminates, there are $k$ sub-networks\footnote{This simple algorithm does not guarantee there will be no overlap between sub-networks, however, the constant time idea is relevant for large networks, where the network size comes to our advantage. An overlap is possible, but the probability decreases with the network size.} of size $M$, and for each network we need to find one target node. 
The best node in each sub-network can be found with an arbitrarily small error 
by iterating through the sub-network nodes and using Monte-Carlo sampling in each sub-network separately. The total running time\footnote{See \cite{v011a004} for the proof of the Monte-Carlo sampling time. As in their experiments, the actual sampling time needed is much shorter than the proof requires, and in practice linear sampling time yields same results as the quadratic time required for the proof.} is $\mathcal{O}(k\cdot M^3)$, which is independent of the network size.

This is different from the algorithms suggested by \cite{Borgs:2014:MSI:2634074.2634144,Wang2012,PhysRevX.4.021024} 
in two main aspects: first, the local sub-networks are not restricted to local trees, but may include cycles, as long as they are within the predetermined sub-network size. This may  allow us to capture effects of highly connected local cliques. 
Second, and more importantly, the running time of random-local-optimization is independent of the network size; 
local sub-networks are not constructed around all (or a fraction of) the nodes in the network, but only around $k$ nodes. The algorithm does not need to know the whole network, nor its size. 

As we show next,  
this simple and fast algorithm 
gives similar results to standard optimization in terms of the influenced-set size in both synthetic and real-world networks, and considering the optimization costs and uncertainty of the system parameters, our random-local-optimization algorithm yields higher expected utility. 
 
%
  
First, we look at the influenced-set size: Figure \ref{fig:HC_and_random_local_both} shows the median influence gained by random-local-optimization and by standard optimization using hill-climbing, as percentages of the network size, for ER networks of size 10,000 (Figure \ref{fig:HC_and_random_local}) and for the High Energy Physics Theory network (Figure \ref{fig:HC_and_random_local_empirical}). It can be seen that in both cases random-local-optimization gives similar results to standard optimization in most of the parameter space, with a modest advantage to hill-climbing only near the critical point. These results were robust: In all synthetic network models and real-world networks we have tested, our algorithm managed to capture similar gains to hill-climbing for most of the parameter range.\footnote{See Table \ref{M_values_table} in the Appendix for robustness to different $M$ values.}

Percolation theory can give some intuition for understanding the empirical success of the random-local-optimization algorithm, by looking at the high and low phases (see, e.g., \cite{gandolfi1992uniqueness,molloy1998size,aharony2003introduction,grimmett1999percolation}). 
At the low phase, there are almost surely no clusters that scale as any power of the system size (in the limit of infinite volume indeed there are only finite clusters at the low phase). 
In this regime, at first approximation the most that can be achieved by using optimization is related to the degree of the chosen seeds, and in typical networks with short linear distances between nodes, a local search is sufficient for finding high degree nodes. 

At the high phase, in every realization of the process there is a single giant cluster, while all other clusters typically scale at most logarithmically with the network size. In this regime, the purpose of optimization is to increase the probability of hitting the giant cluster. The basic idea is that the giant cluster is always close to anywhere in the network. If $M$ is larger than the logarithm of the system size (say, any poly-logarithm with a degree above $1$), then every sub-network of size $M$ will contain in every realization a significant fraction of connected giant-cluster nodes (for this realization). These nodes are easily found in the local optimization process, and although they are not guaranteed to be the same nodes, or to have the same influence as the nodes found in a global greedy optimization process, empirically they are indeed different nodes, but with similar influence distributions. 
A possible rule of thumb for choosing $M$ would be to take the natural logarithm of the network size to a power grater than $1$ (say, $1.5$); however, since the largest relevant empirical networks are no more than an order of magnitude of $10^{10}$ nodes, for which the natural logarithm is $23$, we suggest that a choice of $M=100$ should be sufficient for all networks -- as larger values are not expected to give any substantial improvement. 
We were not yet able to provide formal proofs for the success of the random-local-optimization algorithm, but we found it robust across all the networks we have tested. It would be interesting to try and formulate proofs for specific network models. 




\begin{figure}[t!]
\begin{subfigure}{.49\linewidth}
  \includegraphics[width=1.10\linewidth]{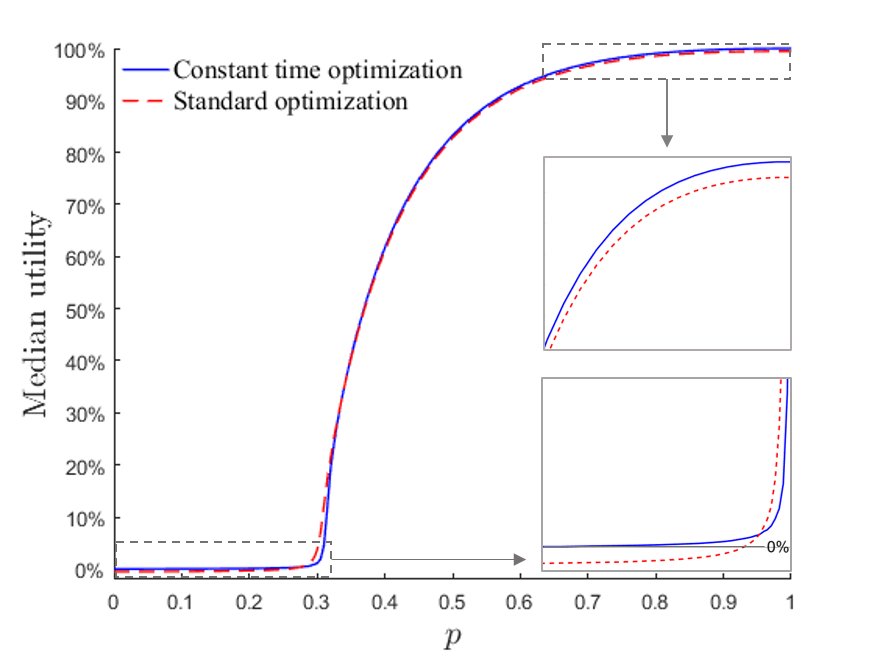}
  \caption{ }  
  \label{fig:opt_and_local_Utility_N5000}
\end{subfigure}
\begin{subfigure}{.49\linewidth} 
  \includegraphics[width=1.10\linewidth]{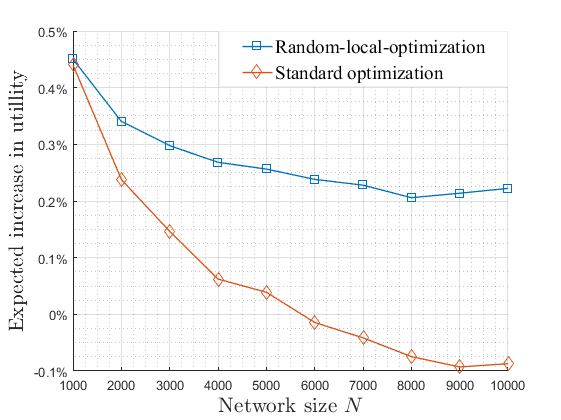}
  \caption{ }
  \label{fig:Avg_over_prior_local_vs_opt}
\end{subfigure}%
\caption{Performance of random-local-optimization and standard optimization in terms of utility. \textbf{(a)} Median utilities of standard and constant-time optimization as percentages of the total value on ER networks of size $N=10$,$000$. The dashed red line is the hill-climbing result, after deducting an optimization cost of $N{}log(N)$ running time of state-of-the-art methods. The blue line is the random-local-optimization result, using sub-networks of size $M=100$, after deducting its constant running time cost. The insets show zoom-in images at high and low probabilities. 
	\textbf{(b)} Expectation over a uniform prior of the parameter $p$ of the increase in utility above the random benchmark, for standard optimization using results of hill-climbing with the lower running time of $N{}log(N)$, and of constant time optimization using random-local-optimization, as a function of $N$, in percentages of the total value.}
\label{fig:utility_figures}
\end{figure}

%
%

Next we turn to demonstrate the implications of these results for the utility measure. Figure \ref{fig:opt_and_local_Utility_N5000} shows the median utility gained by using random-local-optimization and by standard optimization (using hill climbing) on ER networks of size $N=$10,000, after deducting the optimization time costs with a relative cost of $C/v=10^{-3}$ per time unit. The results are presented as percentages of the total value in the network, $v \cdot N$. For the costs of standard optimization we considered the running time of $N{}log(N)$ of state-of-the-art methods, which are faster than hill-climbing, but have lower or similar performance guarantees. 
As discussed in Section \ref{subsection:Considering optimization costs}, since the total cost is the product $C\cdot T$, choosing a smaller value for the constant $C/v$ is equivalent to taking a longer running time. In the standard optimization methods this will be equivalent to measuring the performance on larger networks.  

As can be seen, the utility gained by random-local-optimization is similar to standard optimization for most values of $p$. 
A closer look shows (see the zoom-in insets in Figure \ref{fig:opt_and_local_Utility_N5000}) that our algorithm achieves the goals of the constant time optimization approach: $(i)$ it gives higher utility for most of the parameter range; $(ii)$ near the critical point it is not much worse than hill-climbing. We also see that at the low phase, i.e., at low probabilities, standard optimization may result in negative utility. 
This is because the optimization costs remain high (of the order of the network size), even when the added value is small.

Finally, we compare the performance of random-local-optimization and of standard optimization in a situation of uncertainty of the influence parameters. Since the true value of $p$ is unknown, we have some prior distribution $f(p)$ regarding its value, and specifically in the following, having no other information, we consider a uniform prior. 
Figure \ref{fig:Avg_over_prior_local_vs_opt} depicts the expected added utilities\footnote{This is done by interpolating the data points of the added utilities and calculating the integrals over the uniform prior of $p$ numerically.} 
(again, in respect to the random benchmark) 
over a uniform prior of the values of $p$, $f(p) \sim \mathcal{U}(0,1)$, 
as a function of the network size, presented as percentages of the total value in the network. 
We can see that for this natural situation where $p$ is unknown, our local optimization manages to give higher utility than standard optimization, with a gap that increases with the network size. 

\section{Conclusion}
The influence maximization problem and the related algorithms focus on the structural aspect of network influence, and search 
for seed-sets that have some structural advantage in the network. 
In this study we have found that such structural differences between seed-sets become significant only near a phase transition of the network, in a parameter region that shrinks as a power law of the network size, while away from the phase transition most seed-sets give similar results. 
One implication we have shown is that when there are optimization costs, optimizing over the whole network is profitable only in a narrow parameter region, 
while usually the decision whether or not to invest the optimization costs is being made without knowing the actual parameters. 
This implies that for the structural aspect of optimization, sub-linear algorithms are more attractive and profitable, and we have demonstrated one example in this vein by the random-local-optimization algorithm, which maintains constant optimization time and is effective for a wide parameter range. 
It is interesting 
to further refine this approach, and to design improved sub-linear algorithms that are robust to different parameter configurations. 

The findings of this work also emphasize the importance of studying aspects of influence maximization which are not purely network-structural, but are related to the system's parameters: Which nodes are more susceptible? 
What are the efficient methods to measure, map and predict the spreading parameters in real-world networks? How do the parameters change with time? Is there a dynamic coupling between the (local) spreading probability and the (global) state of network influence?  
Of course, every spreading model that relies only on short-range interactions provides a partial description of reality, as long-range interactions, global fields, or other effects may also play an important role in real-world dynamics \cite{goel2015structural}. Models with such global effects may lead to richer phase maps (e.g., with first order transitions, or co-existence of phases), and it will be interesting to study such models and evaluate them over real-world datasets. 
In addition, our results show that it is important to study the problem of influence maximization across a wide sample of the parameter space, as the results and conclusions from experiments are dramatically different for different parameters. 

Finally, the relation we have found between a physical phase transition of the network and the possible gains from optimization may be more general. 
The results of the present study raise a generic conjecture that there is a class of computationally hard optimization problems that are 
empirically hard only in a limited (zero-measure) range of parameters, 
which is related to phase transitions of the underlying systems. 
The 
characterization of this class of problems and the 
understanding of the relation between computational complexity and phase transitions pose 
a challenge for future research. 
        

\subsubsection*{Acknowledgments\\}
We would like to thank Noam Nisan for discussions and advice throughout this research. 
We also thank Yaron Singer and Gali Noti for helpful conversations, and Dietrich Stauffer, Roi Reichart 
and Effi Levi for comments on an earlier draft.  
This project has received funding from the European Research Council (ERC) under the European Union’s Horizon 2020 research and innovation programme (grant agreement No 740282).

\appendix

\subsection*{APPENDICES}\label{sec:apdx}

\section{Different $k$ Values}\label{subsection:k_vals}
In Section \ref{sec:When_Do_We_Need} we have seen that only near the phase transition there is a parameter range 
where optimization is significantly better than a random choice of the seed set  
and beyond this region a random choice gives similar results to the best optimization methods, 
and that the optimization region shrinks as a power law of the network size. 
This result was persistent in our experiments regardless of the seed-set size $k$ (we tried up to $k=100$). 
In this section we focus on the behavior beyond the optimization region. We show that by using results from percolation theory it is possible to understand the 
mechanism in which the seeding process, and the size $k$ of the seed set, affect the 
influence outcome of a random choice outside the optimization region, and to derive predictions that can be confronted with experiments. 

At the low phase, we know from percolation theory that there are only small clusters which scale logarithmically with $N$, so practically, for real-world networks they will be of order $1$. 
Using this result, which is generally valid for random networks, and has been empirically validated for many real-world networks, the leading order prediction for the low-phase regime is simple: each additional seed hits a new cluster and the influence of a random choice -- or of any other method -- is of order $k \frac{log(N)}{N}$ after proper rescaling such that full influence equals $1$, or $k \cdot \mathcal{O}(1)$ without rescaling. 

At the high phase, there is a single cluster of order $N$, and the other clusters scale as $log(N)$. 
The normalized size of the largest cluster is denoted as $S$. 
In this phase the main contribution for the influence at each 
realization is from a seed that hits this cluster. 
The influence is, to a leading order in $\frac{1}{N}$, proportional to the probability to hit this cluster, which is $1-(1-S)^k$. 
In addition, on average, $k(1-S)$ of the seeds will only hit the small clusters, adding a correction term of order $\frac{log(N)}{N}$, that vanishes for large systems.

\begin{figure}[t!]
\centering
\includegraphics[width=0.73\linewidth]{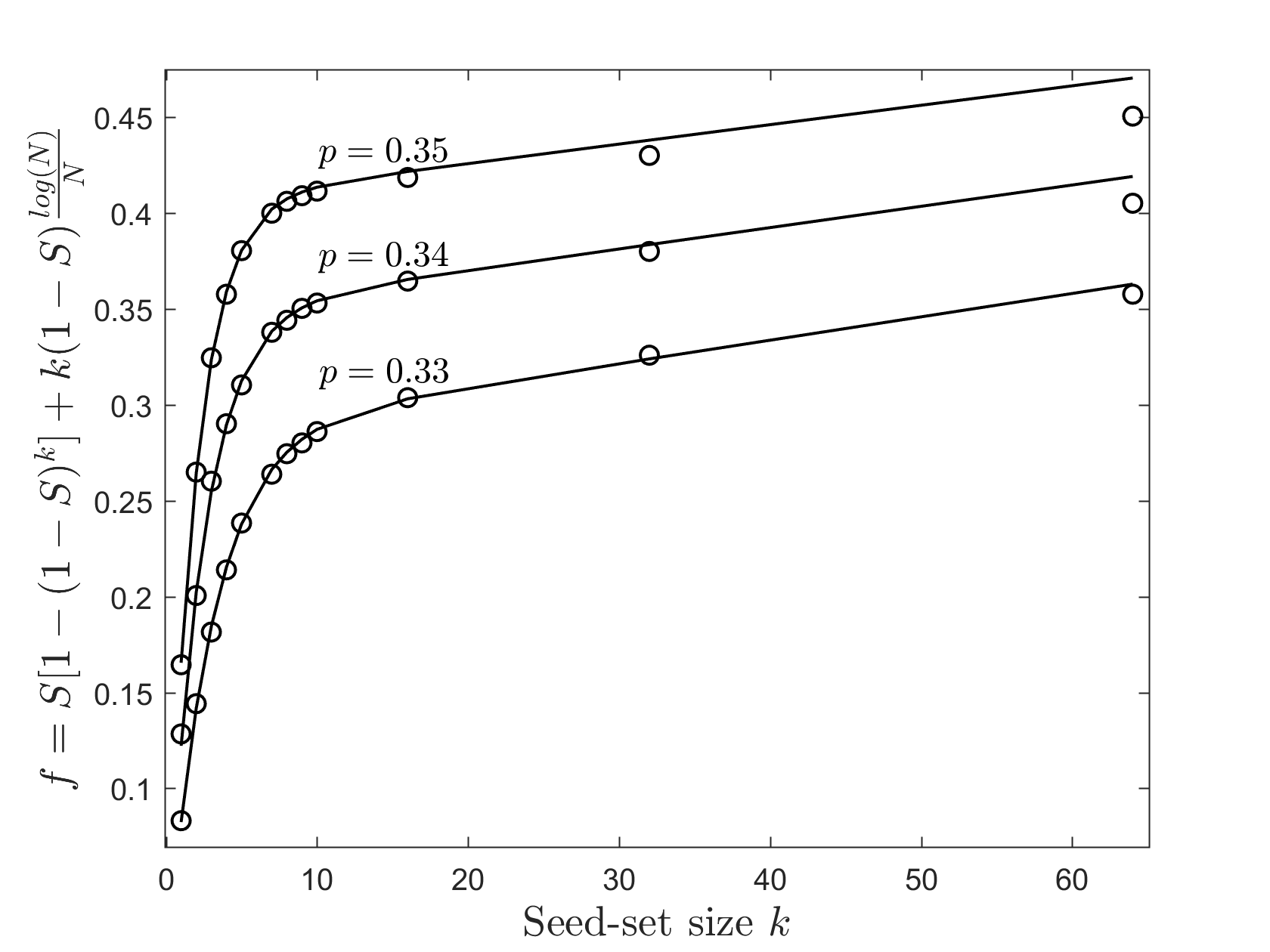}
\vspace{-7pt}
\caption{Effect of different sizes, $k$, of the seed set on the influence obtained by a random choice. The lines are the theoretical predictions for three values of $p$ above the phase transition, and the markers show numerical-experiment results.}
\label{fig:k_vals}
\vspace{-9pt}
\end{figure}

All in all, percolation theory provides 
a clear prediction, that the average influence $f$ depends on 
$k$  
according to:
\vspace{-5pt}
\begin{equation}
f = k \cdot \mathcal{O}(\frac{log(N)}{N})
\label{eqn3}
\end{equation}
\vspace{-6pt}
at the low phase, and according to:
\vspace{-2pt}
\begin{equation}
f = S\left[1- (1-S)^k\right] + k(1-S)\frac{log(N)}{N}
\label{eqn2}
\end{equation}
at the high phase.
The results of experiments at the low phase were indeed linear with respect to $k$, and were of order $\frac{log(N)}{N}$, in agreement with the simple prediction of eq.\ref{eqn3}. The more interesting result of eq.\ref{eqn2} is presented in 
Figure \ref{fig:k_vals} that demonstrates a comparison of this theoretical prediction with numerical experiments on ER networks. 
%
%
\vspace{-3pt}

Because $k$ appears here as an exponent of $(1-S)$, the expression is saturated fast even for small seed sets, and away from the phase transition a random choice almost always hits the largest cluster. 
High values of $k$ become even more redundant from the optimization perspective as they start hitting the same clusters, and a negative correction to the influence of order $\left( k log(N)/N\right)^2$  
becomes significant in moderate size systems. 
This second correction, which is excluded from Figure \ref{fig:k_vals}, explains the overshooting seen in the figure for the higher values of $k$. 
\begin{table} [t!]   
\begin{center} 
 \begin{tabular}{|c c c c c|} 
 \hline 
 $\rule{0pt}{2.5ex}  M:$ & 25 & 50 & 100 & 200 \\ [0.3ex] 
 \hline
 GrQC:          &  0.9841  &  0.9863  &  0.9894  &  0.9900  \\
 \hline
 HEP-Th:        &  0.9932  &  0.9947  &  0.9954  &  0.9958  \\ 
 \hline
 Cond-Mat:      &  0.9969  &  0.9981  &  0.9986  &  0.9988  \\
 \hline
 HEP-Ph:        &  0.9967  &  0.9973  &  0.9978  &  0.9978  \\
 \hline
 Astro-Phys: 	  &  0.9982  &  0.9988  &  0.9989  &  0.9989  \\
 \hline
 ER, N=10,000:  &  0.9947  &  0.9949  &  0.9953  &  0.9955  \\ 
 \hline
 SW, N=10,000:  &  0.9925  &  0.9928  &  0.9932  &  0.9934  \\  
 \hline
 CM, N=10,000:  &  0.9913  &  0.9933  &  0.9946  &  0.9957  \\   
 \hline 
\end{tabular} 
\vspace{1pt} 
\caption{Robustness random-local-optimization to $M$ values: The entries in the table are the ratios of the expected empirical performances of random-local-optimization and hill-climbing for different values of the constant $M$, i.e., $\left(\int^{1}_{0}local_M(p) dp\right)/\left(\int^{1}_{0}opt(p) dp\right)$ over a uniform prior distribution of the parameter $p$, for the empirical network dataset and for ER, SW and CM networks of size $N=10$,$000$.} 
\label{M_values_table}
\end{center}
\end{table} 

\vspace{-8pt}

\section{Random Local Optimization: Different $M$ Values}
In Section \ref{section:Constant time optimization} we have seen the performance of the random local optimization method using our rule of thumb for choosing the mass, $M$, of the sub network size. 
Table \ref{M_values_table} demonstrates the robustness of the performance also to different choices of $M$. The values in the table are the ratios between the cumulative marginal gain over the values of $p$ of the global optimization result (using hill-climbing) and of the random-local-optimization result for the five empirical networks and for ER, SW and CM networks of size $N=$10,000. 

\bibliographystyle{splncs03}
\bibliography{influence_project}

\end{document}